\documentclass[12pt]{article}
\usepackage{fullpage}
\title{    q-DEFORMED HARMONIC OSCILLATOR IN PHASE SPACE}
\author{
\bf A.K.Aringazin$^{\dagger\ddagger}$, K.M.Aringazin$^{\dagger}$,
\bf S.Baskoutas$^{*}$,\\
\bf G.Brodimas$^{*}$, A.Jannussis$^{*\ddagger}$ and E.Vlachos$^{*}$ \\[1cm]
     $^{\dagger}$Department of Theoretical Physics,
                 Karaganda State University,\\
                 Karaganda 470074, Kazakhstan\\
            $^*$ Department of Physics,
                  University of Patras,\\
                  Patras 26110, Greece\\
    $^{\ddagger}$The Institute for Basic Research,
                 P.O. Box 1577,\\
                 Palm Harbor, FL 34682, U.S.A.}

\date{{\it Proc. Intern. Conf. "Advances in Fundamental Physics", Olympia,
Greece, 27-30 Sept. 1993, Eds. M.Barone and F.Selleri, Hadronic Press, 1995,
pp. 329-348}}

\begin{document}
\begin{titlepage}
\maketitle
\abstract{Relation between Bopp-Kubo formulation and Weyl-Wigner-Moyal
symbol calculus, and non-commutative geometry interpretation
of the phase space representation of quantum mechanics
are studied.
 Harmonic oscillator in phase space via creation and annihilation
operators, both the usual and $q$-deformed,
is investigated.
 We found that the Bopp-Kubo formulation is just non-commuting
coordinates representation of the symbol
calculus.
 The Wigner operator for the $q$-deformed harmonic oscillator is
shown to be proportional to the 3-axis spherical angular momentum
operator of the algebra $su_{q}(2)$.
 The relation of the Fock space for the harmonic oscillator and
double Hilbert space of the Gelfand-Naimark-Segal construction is
established.
 The quantum extension of the classical ergodiicity condition is
proposed.}
\end{titlepage}

\setcounter{page}{2}
\section{Introduction}

Phase space formulation of quantum mechanics since pioneering
papers by Weyl\cite{Weyl-27}, Wigner\cite{Wigner-32} and
Moyal\cite{Moyal-49} who were motivated by the obvious
reason to realize quantum mechanics as some extended
version of the Hamiltonian mechanics rather than somewhat
sharp step to the theory of operators acting on Hilbert space,
now becomes of special interest at least for two major reasons.

The first reason is that the quantum mechanics in phase space
represents an example of theory with non-commutative geometry.
Moyal\cite{Moyal-49} and Bayen {\it et al.}\cite{Bayen-78}
developed non-commutative algebra of functions on phase space
which is aimed to represent non-commutative property of
the operators. In turn, the operators are sent to functions
on phase space - symbols - due to the symbol
map\cite{Weyl-27,Wigner-32}, which is well defined
one-to-one map. So, the symbol calculus\cite{Hormander-79,Berezin-80}
provides a reformulation of whole machinery of quantum
mechanics in terms of non-commutative functions on phase space.

Also, Bopp\cite{Bopp-61} and Kubo\cite{Kubo-64} extended the phase
space and introduced non-commutative variables in terms
of which they expressed the Wigner operator\cite{Wigner-32,JP-75}
and the Wigner density operator\cite{Kubo-64,Feynman-72,BJP-76}.
The Bopp-Kubo formulation deals with functions on the extended
phase space which are also non-commuting due to the non-commutative
character of the variables.

Recent studies of the non-commutative phase
space\cite{Wess-90}-\cite{Dimakis-92}
%%%%FULL:\cite{Wess-90,Bertocci-91,Dimakis-92}
are much in the spirit of modern non-commutative
geometry\cite{Connes-85}-\cite{Coquereaux-92}.
%%%%FULL: \cite{Connes-85,Manin-88,Dubois-89,Connes-90,Coquereaux-92}.
Exterior differential calculus in the quantum mechanics in phase
space has been proposed recently by Gozzi and Reuter\cite{GR-93,GR-93a}.
They studied in detail algebraic properties of the symbol
calculus, and have found\cite{GR-93b}, particularly,
quantum analogue of the classical canonical transformations.
Gozzi and Reuter have argued that the quantum mechanics in phase
space can be thought of as a smooth deformation of the
classical one.

Jannussis, Patargias and Brodimas\cite{JPB-78} have constructed
creation and annihilation operators in phase space, and studied
harmonic oscillator in phase space\cite{JP-77}.
Various problems related to the Wigner operator, Wigner
distribution function, and the density matrix in phase space
have been investigated in a series of papers
by Jannussis {\it et al.}\cite{GJP-77}-\cite{JLPFFV-82}.
%%FULL:\cite{JP-75,BJP-76,GJP-77,JSPSV-77,JP-78,JFF-80,JLPFFSV-82,JLPFFV-82}.

The second reason of the importance of the quantum mechanics
in phase space is that the resulting formalism is very similar
to the Hamiltonian formulation of classical
mechanics (not surprise certainly).

 An obvious advantage of the phase space formulation of
quantum mechanics is that it arises to a tempting possibility
to exploit this formal similarity, provided by the smooth deformation,
to extend some of the useful notions and tools,
such as action-angle variables, ergodicity, mixing,
Kolmogorov-Sinai entropy, and chaos, which had been elaborated
in Hamiltonian mechanics to quantum mechanics.
The only thing that one should keep in mind here is that
the phase space quantum mechanics deals with the non-commutative
symplectic geometry rather than the usual symplectic geometry.
So, one should take care of this, primarily because the usual
notion of phase space points is lost in non-commutative case so
that one is forced to work mostly in algebraic terms rather than
to invoke to geometrical intuition. For example, it is not obvious
what is an analogue of the Lyapunov exponents when there are no
classical trajectories.

However, as a probe in this direction, we attempt to
formulate, in this paper, the
extention of the classical ergodicity condition.

We should emphasize here that, clearly, it is highly suitable
to have at disposal the phase space formulation before going
into details of quantum mechanical analogues of the
classical chaos and related phenomena.

As to chaos in dynamical sytems, it should be noted that
the evolution equations, both in the classical and quantum
mechanics in phase space, are Hamiltonian flows, which are
deterministic in the sense that there are no source terms
of stochasticity. In view of this, chaos can be still
thought of as an extreme sensitivity of the long-time
behavior of the probability density and, therefore, of the other
observables of interest, to initial state.
Another fundamental aspect of this consideration is the
process of measurements. However, we shall not
discuss this problem here.

As a specific example of quantum mechanical system in
phase space, we consider, in this paper, one-dimensional
harmonic oscillator.

 We study also the {\it $q$-deformed} oscillator in phase
space which is now of special interest in view of the
developments of quantum algebras\cite{Drinfeld-86}-\cite{Bernard-90}.
%%FULL:\cite{Drinfeld-86,Jimbo-86,Faddeev87,Woronowicz-89,Bernard-90}.
 We should note here that the quantum algebras are particular
cases of the Lie-admissible algebras\cite{J-91}-\cite{JBB-92}.
 The algebra underlying the properties of the $q$-oscillator
in phase space appears to be
the algebra $su_{q}(2)$\cite{Biedenharn-89}-\cite{Fiore-93}.
%%FULL:\cite{Biedenharn-89,Macfarlein-89,Kulish-90,Kachurik-90,SK-93,Fiore-93}.

 The paper is organized as follows.

 In Sec 2.1, we briefly recall the Bopp-Kubo
formulation of quantum mechanics in phase space.

 Sec 2.2 is devoted to Weyl-Wigner-Moyal symbol-calculus
approach to quantum mechanics the main results of which
are sketched.

 In Sec 2.3, we discuss, following Gozzi and Reuter\cite{GR-93b},
modular conjugation and unitary transformations.
 We show that the Bopp-Kubo formulation and the symbol calculus
are explicitly related to each other.
 We give an interpretation of the quantum mechanics in phase
space in terms of non-commutative geometry.
 Quantum mechanical extention of the classical ergodicity
condition is proposed.

 In Sec 2.4, we study translation operators in phase space.
Commutation relations of the Bopp-Kubo translation operators,
both in Hamiltonian and Birkhoffian cases, are presented.

 The results presented in Sec 2 are used in Sec 3 to study
the harmonic ($q-$)oscillator in phase space.

 In Sec 3.1, we present the main properties of the
one-dimensional oscillator in terms of annihilation and
creation operators in phase space. We identify fundamental
$2D$-lattice structure of the phase space resulting from the
commutation relations of the Bopp-Kubo translation operators.
The Fock space for the oscillator is found to be related
to the double Hilbert space of the Gelfand-Naimark-Segal
construction.

 In Sec 3.2, we study $q$-deformed harmonic oscillator in
phase space. The Wigner operator is found to be proportional
to the 3-axis spherical angular momentum operator of the
algebra $su_{q}(2)$.
 Also, the Wigner density operator appeared to be related to
the 3-axis hyperbolical angular momentum operator of the
algebra $su_{q}(1,1) \approx sp_{q}(2,R)$.

\section{Phase space formulation of quantum mechanics}

\subsection{Bopp-Kubo formulation. Non-commutative coordinates}

In studying Wigner representation\cite{Wigner-32}
of quantum mechanics, Bopp\cite{Bopp-61} and Kubo\cite{Kubo-64}
started from classical Hamiltonian $H(p,q)$ and used the
variables (see also \cite{JP-78,JPLFFSV-82})

\begin{eqnarray}
                                                     \label{PQ}
P= p - \frac{i\hbar}{2}\frac{\partial}{\partial q},\quad
Q= q + \frac{i\hbar}{2}\frac{\partial}{\partial p} \\    \label{PQ*}
P^*= p + \frac{i\hbar}{2}\frac{\partial}{\partial q},\quad
Q^*= q - \frac{i\hbar}{2}\frac{\partial}{\partial p}
\end{eqnarray}
instead of the usual $(p,q)$
and obtained the Wigner operator, $W_-$, and the Wigner density operator,
$W_+$, in the following form:
\begin{equation}
W_{\pm} = H(P,Q) \pm H(P^*,Q^*)                     \label{W}
\end{equation}
These operators enter respectively the Wigner equation
\begin{equation}
i\hbar\partial_t \rho =  W_- \rho                    \label{Wigner}
\end{equation}
and the Bloch-Wigner equation\cite{Kubo-64}
\begin{equation}
\partial_\beta F + \frac{1}{2} W_+ F = 0 \qquad\qquad
                      \beta = \frac{1}{kT}         \label{BlochWigner}
\end{equation}
Here, $\rho=\rho (p,q)$ is the Wigner distribution
function\cite{Wigner-32,JLPFFV-82}
and $F=F(p,q,p',q';\beta )$ is the Wigner density matrix\cite{BJP-76}.

As it is wellknown, the Wigner equation (\ref{Wigner}) is
a phase space counterpart of the usual von Neumann equation of quantum
mechanics while the Bloch-Wigner equation (\ref{BlochWigner})
describes quantum statistics in phase space\cite{Feynman-72,GJP-77}.

In view of the definitions (\ref{PQ})-(\ref{PQ*}) of the
variables, it is quite natural to
treat the above formulation in terms of non-commutative
geometry\cite{Connes-90}.

With the usual notation,
$\phi^i = (p_{1} , \dots , p_n ,q^1 , \dots , q^n )$,
$\phi^i \in M_{2n}$,
the first step is to extend the phase space $M_{2n}$
to the (co-)tangent phase space $TM_{2n}$ and define the complex
coordinates,
\begin{equation}
\Phi^{i}_{\pm} = \phi^i \pm
   \frac{i\hbar}{2}\omega^{ij}\frac{\partial}{\partial \phi^{j}}
                                                            \label{Phi}
\end{equation}
where $\omega^{ij}$ is a fundamental symplectic tensor,
$\omega^{ij}=-\omega^{ji}; \ \omega_{ij}\omega^{jk}=\delta^{k}_{i}$.

We observe immediately that these coordinates are non-commutative,
\begin{equation}
\bigl[\Phi^{i}_{\pm},\Phi^{j}_{\pm}\bigr] = \pm i\hbar\omega^{ij}
\qquad   \bigl[\Phi^{i}_{\pm},\Phi^{j}_{\mp}\bigr] = 0  \label{comff}
\end{equation}
and do not mix under time evolution.
The natural projection $TM_{2n} \rightarrow M_{2n}$ comes
with the classical limit $\hbar \rightarrow 0$.

Commutation relations (\ref{comff}) imply that the
"holomorphic" functions, $f(\Phi_-)$, and "anti-holomorphic"
functions, $f(\Phi_+)$, form two mutually commuting closed
algebras on space of functions $C(TM_{2n})$.

Thus, the holomorphic, $H(\Phi_{-})$, and anti-holomorphic,
$H(\Phi_{+})$, Hamiltonians define two separate dynamics,
which are not mixed.
Wigner operators (\ref{W}) are simply sum and difference
between these two Hamiltonians, respectively,
\begin{equation}
W_{\pm} = H(\Phi_- )  \pm H(\Phi_+ )               \label{WPhi}
\end{equation}
 So, physical dynamics comes with the combinations of these
two Hamiltonians. In the classical limit, the Wigner operators
cover the Liouvillian $L$ and the Hamiltonian,
\begin{equation}
W_{-} = -i\hbar L + O(\hbar^2 )  \qquad           \label{Wclass}
W_{+} = 2H(p,q) + O(\hbar^2 )
\end{equation}
where $L \equiv \ell_{h}=-h^{i}\partial_{i}$ is a Lie derivative along
the Hamiltonian vector
field $h^{i} = \omega^{ij}\partial_{j}H$\cite{Arnold-78,Abraham-78}.

According to complex character of the variables (\ref{Phi}),
one can define the involution ${\cal J}$ acting simply
as complex conjugation
\begin{equation}
{\cal J}:\ \Phi^{i}_{\pm} \rightarrow \Phi^{i}_{\mp}           \label{JPhi}
\end{equation}
This involution may be thought of as a conjugation
interchanging the two pieces of the physical dynamics.

\subsection{Weyl-Wigner-Moyal formulation. Symbol calculus}

In order to achieve phase space formulation of quantum mechanics,
Weyl\cite{Weyl-27} and Wigner\cite{Wigner-32} introduced
symbol map associating with each operator $\hat A$, acting on
Hilbert space, a symbol $A(\phi )$, function on phase space,
$A(\phi ) = symb(\hat{A})$, due to
\begin{equation}
A(\phi ) =
\int \frac{d^{2n}\phi_{0}}{(2\pi\hbar)^n}
     exp\Bigl[\frac{i}{\hbar}\phi^{i}_{0}\omega_{ij}\phi^{j}\Bigr]
     Tr\bigl( \hat T (\phi_{0})\hat A \bigr)                 \label{symbol}
\end{equation}
with
\begin{equation}
\hat T (\phi_{0}) =
     exp\Bigl[\frac{i}{\hbar}\phi^{i}_{0}\omega_{ij}\hat\phi^{j}\Bigr]
                                                           \label{TWeyl}
\end{equation}
The symbol map is well defined invertible one-to-one map from space
of operators, ${\cal O}$, to space of functions depending on phase space
coordinates\cite{Hormander-79,GR-93a},
${\cal O} \rightarrow C(M_{2n})$.
Particularly, Hermitean operators are mapped to real functions,
and vice versa.

The key property of the symbol calculus is that the ordinary
pointwise product of the functions is appropriately generalized
to reproduce the non-commutative product of the operators.
The product on $C(M_{2n})$, making the symbol map an
algebraic homomorphism, is the Moyal product\cite{Moyal-49,Berezin-80},
\begin{eqnarray}
                                                     \label{mp}
(A*B)(\phi ) &=&
symb(\hat A \hat B) \nonumber \\
             &=&
A(\phi )
exp\bigl[\frac{i}{2\hbar}\bar\partial_{i}\omega_{ij}\vec\partial_{j}\bigr]
B(\phi ) \\
            &=&
A(\phi )B(\phi ) + O(\hbar ) \nonumber
\end{eqnarray}
The Moyal product is associative but apparently non-commutative,
and represents, in $C(M_{2n})$, non-commutative property of the algebra of
operators,
and non-local character of quantum mechanics.

The Moyal bracket\cite{Moyal-49}
\begin{eqnarray}
\{ A, B \}_{mb} &=&                             \label{mb}
symb(\frac{1}{i\hbar}\bigl[ A, B ] ), \nonumber \\
                  &=&
\frac{1}{i\hbar}( A*B - B*A) \\
                  &=&
\{ A, B \}_{pb} + O(\hbar^2)             \nonumber
\end{eqnarray}
is a symbol of commutator between two operators, and reduces to
the usual Poisson bracket $\{ .,. \}_{pb}$ in the
classical limit.
 Thus, the algebra $(C(M_{2n}), \{ ,\}_{mb})$ is an algebra
of quantum observables, and it can be continuously
reduced to the algebra $(C(M_{2n}), \{ ,\}_{pb})$ of classical
observables.

Symbol map of the von Neumann's equation is written as\cite{GR-93a}
\begin{eqnarray}
                                                     \label{me}
\partial_t \rho (\phi , t) &=&
-\{ \rho, H \}_{mb} \nonumber \\
                           &=&
-\ell_h \rho + O(\hbar^2 )
\end{eqnarray}

In the classical limit, this equation covers
the Liouville equation of classical mechanics\cite{Koopman-31,Neumann-32}.

To summarize, the symbol calculus can be treated as a smooth deformation
of classical mechanics linking non-associative Poisson-bracket algebra of
classical observables, $A(\phi), \dots ,$ and an
associative commutator algebra of quantum observables, $\hat A , \dots $.
Full details of the symbol calculus may be found in \cite{GR-93a,GR-93b}
and references therein.

\subsection{Chiral symmetry and unitary transformations}

Gozzi and Reuter\cite{GR-93b} have investigated recently the
algebraic properties of the quantum counterpart of the classical
canonical transformations using the symbol-calculus approach
to quantum mechanics.

They found, particularly, that the operators $L_f$  and $R_f$
acting as the left and right multiplication with symbol $f$,
respectively,
\begin{equation}
L_{f}g = f*g \qquad  R_{f}g = g*f                     \label{LR}
\end{equation}
form two mutually commuting closed algebras, ${\cal A}_{L}$
and ${\cal A}_{R}$, (cf. \cite{JSPSV-77})
\begin{equation}
                                                       \label{LL}
\bigl[ L_{f_1} ,L_{f_2} \bigr] = i\hbar L_{\{f_1 f_2\}_{mb} } \qquad
\bigl[ R_{f_1} ,R_{f_2} \bigr] =-i\hbar R_{\{f_1 f_2\}_{mb} } \qquad
\bigl[ L_{f_1} ,R_{f_2} \bigr] = 0
\end{equation}
which are explicitly isomorphic to the original Moyal-bracket
algebra on $C(M_{2n})$.

    Also, $L_f$  and $R_f$ can be presented by virtue of the
Moyal product (\ref{mp})  as\cite{GR-93b}
\begin{equation}
L_{f} = : f(\Phi^{i}_{+}) : \qquad
R_{f} = : f(\Phi^{i}_{-}) :                      \label{LRPhi}
\end{equation}
where $\Phi^{i}_{\pm}$ are defined by (\ref{Phi}), and $:\dots :$
means normal ordering symbol (all derivatives $\partial_{i}$ should be
placed to the right of all $\Phi$'s).

It has been shown\cite{GR-93b} that the linear combinations of the
above operators,
\begin{equation}
V^{\pm}_{f} = \frac{1}{i\hbar}(L_f \pm R_f )             \label{V}
\end{equation}
for real $f$, generate non-unitary,
$\hat g \rightarrow \hat{U}\hat g \hat{U}$,
and unitary,
$\hat g \rightarrow \hat{U}\hat g \hat{U^{-1}}$,
transformations, respectively ($\hat{U}$ is an unitary operator).

We see that the Wigner operators, $W_{\pm}$, given in the
Bopp-Kubo formulation by (\ref{WPhi}), are just
\begin{equation}
W_{\pm}  = L_H \pm R_H  = i\hbar V^{\pm}_{H}        \label{WLR}
\end{equation}
so that the Wigner equation (\ref{Wigner}) can be written as
\begin{equation}
\partial_t \rho =  V^{-}_{H} \rho             \label{Wigner2}
\end{equation}
where $H$ is the Hamiltonian. So, we arrive at the conclusion that the
representations (\ref{LRPhi}) provide the relation between the Bopp-Kubo
and Weyl-Wigner-Moyal formulations.

Various algebraic properties of the generators $V^{-}_{f}$
have been found by Gozzi and
Reuter\cite{GR-93b}. Particularly, they found that
in two dimensional phase space the generators $V^{-}_{f}$, in the basis
$V_{\vec m} = -exp(i\vec m \vec \phi),\
\vec m = (m_1 , m_2) \in Z^2 $, on torus $M_2 = S^1 \times S^1$,
satisfy a kind of the $W_{\infty}$-algebra commutation relations,
\begin{equation}
\bigl[ V^{-}_{\vec m}, V^{-}_{\vec n} \bigr] =
\frac{2}{\hbar}                           \label{Winfty}
sin(\frac{\hbar}{2}m_{i}\omega^{ij}m_{j})V^{-}_{\vec m + \vec n}
\end{equation}
which are deformed version of the $w_{\infty}$-algebra
of the classical $sdiff(T^2 )$, area preserving diffeimorphisms
on the torus.

Also, an important result shown in \cite{GR-93b} is that $V^{-}_{f}$ is
invariant under the modular conjugation operator defined on symbols by
\begin{equation}
{\cal J} f = f^*  \qquad
{\cal J}(f*g)  = {\cal J}(g)*{\cal J}(f)  \label{J}
\end{equation}
Namely,
\begin{equation}
{\cal J} L_f {\cal J} = R_f \qquad                   \label{JLJ}
{\cal J} R_f {\cal J} = L_f \qquad
{\cal J} V^{-}_f{\cal J} = V^{-}_f \qquad
{\cal J} V^{+}_f{\cal J} = -V^{+}_f
\end{equation}
 This symmetry resembles the {\it chiral} symmetry and seems
to be broken in the classical mechanics.
 This argument is supported by the fact that the Moyal
product (\ref{mp}) becomes commutative in the classical limit.
Indeed, in the classical case, the difference between the left
and right multiplications
on $C(M_{2n})$ disappears so that there is no room for the
modular conjugation operator ${\cal J}$, and the original algebra
${\cal A}_{L}\otimes{\cal A}_{R}$ is contracted to its diagonal
subalgebra\cite{GR-93b}.

 The operator $V^{+}_f$ seems to have no analogue in the geometry
of phase space of classical mechanics
since $V^{+}_f$ blows up at $\hbar \rightarrow 0$ due to the
factor $1/i\hbar$ in the definition (\ref{V}).
However, $i\hbar V^{+}_{f}$ is ${\cal J}$-invariant and has
the classical limit
$i\hbar V^{+}_{f} = 2f + O(\hbar^{2})$
so that
$i\hbar V^{+}_{H} = W_{+}= 2H + O(\hbar^{2})$ is simply
two times Hamiltonian.
In the Bloch-Wigner equation (\ref{BlochWigner}),
$i\hbar V^{+}_f$ plays the role of Hamiltonian
defining the density matrix in quantum statistics\cite{Feynman-72}.

 The operator $V^{-}_f$ has an explicit interpretation\cite{GR-93b} as a
{\it quantum deformed Lie derivative along the hamiltonian vector field}
in accordance with (\ref{Wigner2}).
 Furthermore, in quantum mechanics the ${\cal J}$-invariance of
$V^{-}_f$ provides {\it unitary}  time evolution due to the Wigner
equation (\ref{Wigner2}).

 The structure of the Weyl-Wigner-Moyal calculus, which deals with
non-commutative algebra, may be seen in a more refined way from the
non-commutative geometry\cite{Connes-90,Coquereaux-92}
point of view as follows.

 First, recall that usual definition of topological space $M$ is equivalent
to definition of commutative algebra ${\cal A}$ due to the identification
${\cal A} = C(M)$, with the algebra $C(M)$ of continuous complex valued
functions on $M$ (Gelfand correspondence).
 Conversely, $M$ can be understood as the
spectrum of algebra ${\cal A}$, {\it i.e.} points $x \in M$ are irreducible
representations owing to the relation $x[f] = f(x)$ when $f \in {\cal A}$.
Next step is that one is free to assume that the algebra ${\cal A}$ is
non-commutative in general, and then think about a non-commutative
version of the space $M$. Particularly, classical notion of point
$x \in M$ is modified, in non-commutative geometry, due to the
basic relation mentioned above.

 Specific example we will consider for our aims is a non-commutative
vector bundle.
 In classical geometry, sections of a vector bundle $E$ above
a manifold $M$ play, in physical context, the role of matter fields.
Here, an important point to be noted is that the space ${\cal E}$ of the
sections is a bimodule over the algebra ${\cal A} = C(M)$ of the functions
on $M$.
 In the non-commutative case, there are {\it left} and {\it right}
modules over non-commutative algebra ${\cal A}$ instead of the bimodule.
That is, for $\sigma \in {\cal E}$ and $f \in {\cal A}$, $f\sigma$ and
$\sigma f$ are not both made sense as elements of ${\cal E}$.
One may choose, for convenience, the right module, and then
characterize the non-commutative vector bundle as a quotient
of free module $A^{m}$, {\it i.e.} as the (right)
projective module over the algebra ${\cal A}$,
$\, {\cal E} = P{\cal A}^{m}$, for some projector $P, \, P^{2} = P$,
and some $m$.

 In the symbol calculus, we have, obviously, ${\cal E} = {\cal A}$ itself,
where ${\cal A} = (C(M_{2n}), *)$ is the non-commutative algebra endowed
with the Moyal product. The sections are functions on $M_{2n}$ acting
by the left and right multiplications and forming, respectively,
left and right ${\cal A}$-modules. The modular conjugation acts due to
$$ {\cal J}: {\cal A}\otimes {\cal A} \rightarrow {\cal A}\otimes {\cal A}
$$
$$    (L,R) \mapsto (R,L)
$$
and ${\cal E}$ is the quotient,
${\cal E}= {\cal A}\otimes {\cal A} /{\cal J}$.

 The ${\cal A}$-modules to be {\it unital} one has to put
$I_{L}*f = f$
and
$f*I_{R} = f$, $\forall f \in C(M_{2n})$,
with $I_{L,R}$ being the left and right "identity" elements of ${\cal E}$.
Because
${\cal E}={\cal A}$,
we have actually $I_{L}=I_{R} = I \in C(M_{2n})$ so that the above
conditions imply
$$
f*I - I*f = 0 \qquad
f*I + I*f = 2f \qquad
$$
According to definitions (\ref{LR})  and (\ref{WLR}), these
equations can be rewritten as
\begin{equation}
V^{-}_{f}I = 0 \qquad                 \label{unital}
i\hbar V^{+}_{f}I = 2f \quad\qquad \forall f \in C(M_{2n})
\end{equation}
 The question arises as to existence of such unique function $I$
that the both equations (\ref{unital})
are satisfied for any function $f$.
 We observe that in the classical case the last two equations have correct
limits at $\hbar \rightarrow 0$, and are satisfied for any $f$ identically
only if $I(\phi ) = 1$, as it was expected ($1f = f1 = f$).

 The Bopp-Kubo representation provides a realization of representation
space of the algebra ${\cal A}$, with the variables $\Phi^{i}_{\pm}$,
which extends the usual $M_{2n}$ for the non-commutative case.

In the remainder of this section, we will consider the
extention of the classical {\it ergodicity} condition\cite{Arnold-68}.
Quantum mechanical analogue of the classical condition
of ergodicity can be written as
\begin{equation}
V^{-}_{H}\rho = 0                     \label{ergoGozzi}
\end{equation}
due to comparison of (\ref{me}) and (\ref{Wigner2}),
with the solution $\rho$ being non-degenerate, at least at the classical
level.

 In the classical limit, this equation covers the usual equation,
$L\rho = 0$, where $L$ denotes the Liouvillian, whose non-degenerate
eigenfunctions with zero eigenvalues describe ergodic Hamiltonian
systems\cite{Arnold-68}, which are characterized by the only
constant of motion, energy $H$.
As to solutions, recent
studies\cite{GRT-91a}-\cite{Aringazin-93b}
%%FULL:\cite{GRT-91a,GRT-91c,GR-92a,Aringazin-93a,Aringazin-93b}
of the classical ergodicity condition
within the path integral approach to classical mechanics show that
the solution is given specifically by the Gibbs state form.

 The condition (\ref{ergoGozzi}) can be rewritten in the Bopp-Kubo
representation as
\begin{equation}
H(\Phi_{+})\rho(\phi ) = H(\Phi_{-})\rho(\phi )     \label{ergoBopp}
\end{equation}
where we have used the relations (\ref{WPhi}) and (\ref{WLR}),
that means that the holomorphic and anti-holomorphic Hamiltonians
have the same spectrum.
Also, it is remarkable to note that the equation (\ref{ergoGozzi})
is similar to the first equation of (\ref{unital}), with
$f(\phi ) = H(\phi )$ and $I(\phi ) = \rho(\phi )$.

We pause here with the further discussion stating that more
analysis is needed to verify the proposed extention of the
ergodicity condition (\ref{ergoGozzi}) which may be made
elsewhere.

\subsection{Translation operators}

The operator $T(\phi_0)$ defined by (\ref{TWeyl})
and used to represent the Weyl symbol map (\ref{symbol})
has a meaning of the operator of translations in phase space.
Bopp\cite{Bopp-61} has introduced such an operator in
$\Phi^{i}_{\pm}$-variables representation and Jannussis
{\it et al.} \cite{JPB-78} have studied their properties.

Let us define the translation operators, in the Bopp-Kubo
formulation,
\begin{equation}
T_{\pm}(\phi_{0}) =
exp\bigl[ \pm\frac{i}{\hbar}\phi^{i}_{0}\omega_{ij}\Phi^{j}_{\pm}\bigr]
                                                       \label{T}
\end{equation}
where $\Phi^{i}_{\pm}$ are defined by (\ref{Phi}).
It is easy  to verify that due to the fundamental commutation
relations (\ref{comff}) they build up two mutually commuting
algebras,
\begin{eqnarray}
                                                   \label{TT}
\bigl[ T_{\pm}(\phi_{1}),T_{\pm}(\phi_{2})\bigr] =
 \pm 2i\ sin(\frac{1}{\hbar}\phi^{i}_{1}\omega_{ij}\phi^{j}_{2})
T_{\pm}(\phi_{1} + \phi_{2}) \\
\bigl[ T_{\pm}(\phi_{1}),T_{\mp}(\phi_{2})\bigr]  = 0
\end{eqnarray}

In the case of Birkhoffian generalization of Hamiltonian
mechanics\cite{Aringazin-93a}-\cite{GRT-91b}
one supposes that the symplectic 2-form $\omega$ depends on
phase space coordinates, $\omega = \omega (\phi)$, but it is
still non-degenerate and closed, $d\omega=0$. Consistency of the
Birkhoffian mechanics is provided by the Lie-isotopic
construction\cite{Santilli-88}-\cite{Aringazin-mono-91}.
%%%%FULL:\cite{Santilli-88,Aringazin-AAG-90,Aringazin-mono-91}
In this case, the fundamental commutation relations (\ref{comff})
are essentially modified,
\begin{eqnarray}
                                              \label{comffBirk}
\bigl[ \Phi^{i}_{\pm} , \Phi^{j}_{\pm} \bigr] &=&
\pm i\hbar\omega^{ij}
+ (\frac{i\hbar}{2})^2\omega^{mn}\omega^{ij}_{\ \ ,m}\partial_{n} \\
\bigl[ \Phi^{i}_{\pm} , \Phi^{j}_{\mp} \bigr] &=&
\mp (\frac{i\hbar}{2})^2\omega^{mn}\omega^{ij}_{\ \ ,m}\partial_{n} \nonumber
\end{eqnarray}
Consequently, the commutation relations (\ref{TT}) for the
translation operators are also changed.
Tedious calculations show that
\begin{eqnarray}
                                                   \label{TTBirk}
\bigl[ T_{\pm}(\phi_{1}), T_{\pm}(\phi_{2})\bigr]  =
\pm 2i\ sin \Bigl( \frac{1}{\hbar}\phi^{i}_{1}\phi^{j}_{2}
(\omega_{ij} + \frac{1}{2}\omega_{ij,m}\phi^{m}) \Bigr)
T_{\pm}(\phi_{1}+\phi_{2}) \\
\bigl[ T_{\pm}(\phi_{1}),T_{\mp}(\phi_{2})\bigr]  =  \nonumber \\
\pm 2i\ sin \Bigl( \frac{1}{\hbar}\phi^{i}_{1}\phi^{j}_{2}
(\omega_{im,j} - \frac{1}{2}\omega_{ij,m})\phi^{m} \Bigr)
 exp\bigl[\pm \frac{i}{\hbar}
 (\phi^{i}_{1}\omega_{ij}\Phi^{j}_{\pm}
 -\phi^{i}_{2}\omega_{ij}\Phi^{j}_{\mp})\bigr]
\end{eqnarray}
Here, we have used the identity
$\omega^{im}\omega^{jk}_{\ \ ,m} +
 \omega^{jm}\omega^{ki}_{\ \ ,m}  +
 \omega^{km}\omega^{ij}_{\ \ ,m}  = 0$,
and denote
$\omega_{ij,m} = \partial_{m}\omega_{ij}$.
We see that in the Birkhoffian case the holomorphic and
anti-holomorphic functions do not form two mutually
commuting algebras, in contrast to the Hamiltonian case
characterized by $\omega_{ij,m} = 0$.
 Evidently, the Birkhoffian generalization is important
for the case when the symplectic manifold can not be covered
by {\it global} chart with constant symplectic tensor
$\omega_{ij}$. This is, for example, the case of $M_{2n}$ with
a non-trivial topology. However, it should be noted that
the symplectic manifold can be always covered by local
charts with constant $\omega_{ij}$ due to Darboux theorem.

\section{$q$-deformed harmonic oscillator in phase space}

\subsection{Harmonic oscillator in the Bopp-Kubo phase space representation}

Instead of studying the harmonic oscillator in phase space via
the Wigner equation (\ref{Wigner}) it is more convenient
to exploit corresponding creation and annihilation operators
in the phase space.

Jannussis, Patargias and Brodimas\cite{JPB-78} have defined the
following two pairs of the creation and annihilation operators
following the Bopp-Kubo formulation:
\begin{eqnarray}
                                                      \label{a-}
a_{-}     = \frac{1}{\sqrt 2}
\bigl(\sqrt{\frac{m\omega}{\hbar}}Q+i\sqrt{\frac{1}{m\omega\hbar}}P\bigr) \\
a^{+}_{-} = \frac{1}{\sqrt 2}                         \label{a+-}
\bigl(\sqrt{\frac{m\omega}{\hbar}}Q-i\sqrt{\frac{1}{m\omega\hbar}}P\bigr) \\
a_{+}     = \frac{1}{\sqrt 2}                         \label{a+}
\bigl(\sqrt{\frac{m\omega}{\hbar}}Q^*
+i\sqrt{\frac{1}{m\omega\hbar}}P^*\bigr)\\
a^{+}_{+} = \frac{1}{\sqrt 2}                         \label{a++}
\bigl(\sqrt{\frac{m\omega}{\hbar}}Q^* -
i\sqrt{\frac{1}{m\omega\hbar}}P^*\bigr)
\end{eqnarray}
These operators obey the following usual commutation relations:
\begin{equation}
\bigl[a_{\pm}, a^{+}_{\pm}\bigr] = 1 \quad                    \label{aa}
\bigl[a_{\pm}, a^{+}_{\mp}\bigr] =
\bigl[a_{\pm}, a_{\mp}\bigr]     =
\bigl[a^{+}_{\pm}, a^{+}_{\mp}\bigr]  = 0
\end{equation}
The Bopp-Kubo holomorphic and anti-holomorphic Hamiltonians for
the harmonic oscillator then read
\begin{eqnarray}
                                                     \label{H}
H(P,Q) =
         \frac{P^2}{2m} + \frac{m}{2}\omega^2 Q^2
       =
         \hbar\omega (a^{+}_{-}a_{-} + \frac{1}{2}) \\
H(P^* ,Q^* ) =
         \frac{P^{*2}}{2m} + \frac{m}{2}\omega^2 Q^{*2}
       =
         \hbar\omega (a^{+}_{+}a_{+} + \frac{1}{2}) \\
\end{eqnarray}
and the Wigner operator due to (\ref{W}) takes the form
\begin{eqnarray}
                                                   \label{Waa}
W_{-} &=& \hbar\omega (a^{+}_{+}a_{+} - a^{+}_{-}a_{-}) \\
                                                   \label{Wnn}
      &\equiv& \hbar\omega (\hat{n_1} - \hat{n_2})
\end{eqnarray}
In the two-particle Fock space ${\cal F}_{1} \otimes{\cal F}_{2}$
with the basis $|n_1 \, n_2\rangle$, the pairs of operators
(\ref{a-})-(\ref{a++}) act due to
\begin{eqnarray}
                                             \label{aa+-}
a^{+}_{-}|n_1 \, n_2\rangle = \sqrt{n_1 +1}|n_1 +1 \, n_2\rangle \\
                                             \label{aa-}
a^{ }_{-}|n_1 \, n_2\rangle = \sqrt{n_1}   |n_1 -1 \, n_2\rangle \\
                                            \label{aa++}
a^{+}_{+}|n_1 \, n_2\rangle = \sqrt{n_2 +1}|n_1 \, n_2 +1\rangle \\
                                             \label{aa+}
a^{ }_{+}|n_1 \, n_2\rangle = \sqrt{n_2}   |n_1 \, n_2 -1\rangle
\end{eqnarray}
Then, the Wigner operator (\ref{Wnn}) has the following eigenvalues
\begin{equation}
W_{-}|n_1 \, n_2\rangle = (n_1 - n_2)|n_1 \, n_2 \rangle  \label{W-eigen}
\end{equation}
The eigenfunctions of the Wigner operator (\ref{Wnn}) have
the following form\cite{JPB-78,JP-77}:
\begin{equation}
\varphi_{n_1 n_2}(p,q) =                           \label{varphi}
\int dp_0 dq_0\ T_{+}(p_0 , q_0 )\varphi_{0n_2}(p,q)\varphi_{n_1 0}(p,q)
\end{equation}
where
\begin{equation}
\varphi_{n_1 0}  =
\frac{1}{\pi\sqrt{\hbar}}\frac{1}{\sqrt{n_1 !}}
\Bigl(\frac{2m\omega}{\hbar}\Bigr)^{n_1 /2}
\Bigl( q - i\frac{p}{m\omega}\Bigr)^{n_1}
exp\Bigl(- \frac{2H(p,q)}{\hbar\omega}\Bigr)                  \label{varphi1}
\end{equation}
and the same for $\varphi_{0 n_2}$ with the replacement
$n_1 \rightarrow n_2$ in the r.h.s. of (\ref{varphi1}).

The vacuum is characterized by the Gibbs state form
\begin{equation}
\varphi_{00} = \frac{1}{\pi\sqrt{\hbar}}
     exp\Bigl(- \frac{2H(p,q)}{\hbar\omega}\Bigr)   \label{vacuum}
\end{equation}
The action of the Bopp translation operators on the functions
(\ref{varphi1}) can be easily determined, and the result is
\begin{equation}
T_{\pm}(p_0 , q_0 )\varphi_{n_1 0}(p,q)=
exp\Bigl(\pm \frac{i}{\hbar}(p_0 q - q_0 p)\Bigr)
\varphi_{n_1 0}(p+p_0 , q+q_0 )                     \label{Tvarphi}
\end{equation}
The commutators (\ref{TT}) take the form
\begin{equation}
                                                   \label{TT2dim}
\bigl[ T_{\pm}(p_{1},q_{1}), T_{\pm}(p_{2},q_{2})\bigr]  =
\pm 2i\ sin\frac{1}{\hbar}(p_{1}q_{2}-q_{1} p_{2})
T_{\pm}(p_{1}+p_{2},q_{1}+q_{2})
\end{equation}
The translation operators in (\ref{TT2dim}) commute when
\begin{equation}
\frac{1}{\hbar}(p_{1}q_{2} - q_{1}p_{2}) = \pi l
\qquad l \in Z                                  \label{flux}
\end{equation}
This condition is similar to the one of quantization of magnetic
flux for $2D$-electron gas in uniform magnetic field\cite{JPB-78}.

This means that the phase space asquires $2D$-lattice structure with
the basic unit-cell vectors
$\vec \phi_{1} = (p_{1},q_{1})$
and
$\vec \phi_{2} = (p_{2},q_{2})$
obeying (\ref{flux}), i.e.
\begin{equation}
 \vec n \cdot\vec \phi_{1}\times\vec \phi_{2} = l\Psi_{0}
 \qquad \Psi_{0} = \pi\hbar                            \label{aflux}
\end{equation}
The degeneracy of the energy levels of the harmonic oscillator
in the phase space is then related to the lattice structure.
Namely, the representation (\ref{varphi}) of $\varphi_{n_{1}n_{2}}$
means that one "smears" the product
            $\varphi_{n_{1}0}\varphi_{0n_{2}}$
(a "composite state" of two identical systems) over all the phase space.
So, $\varphi_{n_{1}n_{2}}$ remain to be eigenfunctions with the same
eigenvalues under the translations of the form
     $\vec R = N_{1}\vec \phi_{2} + N_{2}\vec \phi_{2},\  N_{1,2}\in Z,$
leaving the lattice invariant. This is a kind of the magnetic group
periodicity\cite{Zak-64}-\cite{Cristofano-91}.
%%%%FULL:\cite{Zak-64,Prange-87,Cristofano-91}.

 To implement the lattice structure of the phase space explicitly
one may start with the vacuum state (\ref{vacuum}), which is
characterized by zero angular momentum, to define four sets of
functions
\begin{equation}
\varphi^{(\alpha )}_{\vec k}(\vec \phi ) =
\sum_{\vec R^{\alpha}} exp(i\vec k \vec R^{\alpha})   \label{varphiR}
T_{+}(\vec R^{\alpha})\varphi_{00}(\vec \phi )
\end{equation}
where
\begin{eqnarray}
                                                     \label{R}
\vec R^{\alpha} = \vec R_{0} + I^{\alpha}_{i}  \qquad
\vec R_{0} = N_{1}\vec \phi_{2} + N_{2}\vec \phi_{2} \qquad
\alpha = 0,1,2,3  \\
       I^{0}_{i}= (0,0)\qquad
       I^{1}_{i}= (1,0)\qquad
       I^{2}_{i}= (0,1)\qquad
       I^{3}_{i}= (1,1) \nonumber
\end{eqnarray}
and the sum is over all four-sets of the $2D$-lattice points.
The unit cell in the definition of each $\vec R^{\alpha}$
has $4l$ flux quanta $\Psi_{0}$ passing through it.

 Gozzi and Reuter\cite{GR-93b} have argued that there is a close
relation between the symbol-calculus formalism and the
Gelfand-Naimark-Segal construction\cite{Thirring-79}.
 In general, the GNS construction is specifically aimed to
define non-commutative measure and topology\cite{Coquereaux-92}.

 The GNS construction provides bra-ket-type averaging, instead of
the usual trace averaging, in the thermo field theory\cite{Umezava-82}
when one deals with {\it mixed} states.  This construction assumes
a double Hilbert space representation of states,
$||\hat A \rangle\rangle =
\sum A_{\alpha\beta}|\alpha\rangle \otimes|\beta\rangle
                           \in {\cal H}\otimes{\cal H}$.
So, particularly, the average of $\hat A$ is given by
    $\langle \hat A \rangle =
      \langle\langle \hat \rho^{1/2}||
          \hat{A} \otimes \hat{I}
      ||\hat \rho^{1/2}\rangle\rangle $,
with $\hat{I}$ being identity operator.
The modular conjugation operator ${\cal J}$  acts on the double
Hilbert space by interchanging the two Hilbert spaces\cite{GR-93b}.

Time evolution of the GNS density is given by
$i\hbar\partial_{t}||\hat \rho^{1/2}\rangle\rangle
         = H^{-}||\hat \rho^{1/2}\rangle\rangle$,
Here, the GNS Hamiltonian
$H^{-} = \hat{H}\otimes I - {\cal J}(\hat{H}\otimes I){\cal J}$
can be evidently associated with the Wigner operator
$W_{-} = i\hbar V^{-}_{H}$.

In view of the analysis of the oscillator in phase space,
the eigenfunctions $\varphi_{n_{1}0}$ and $\varphi_{0n_{2}}$
can be ascribed to the two pieces of the GNS double Hilbert space.
Also, the GNS double Hilbert space is associated to the
double Fock space ${\cal F}_{1}\otimes{\cal F}_{2}$, with the
modular conjugation operator ${\cal J}$ acting on
${\cal F}_{1}\otimes{\cal F}_{2}$ by interchanging the two Fock
spaces.

\subsection{$q$-deformed harmonic oscillator in phase space}

The $q$-deformation of the commutation relations (\ref{aa})
for the Bopp-Kubo creation and annihilation operators
(\ref{a-})-(\ref{a++}) reads
\begin{equation}
b_{-}b^{+}_{-} - \frac{1}{q}b^{+}_{-}b_{-} = q^{\hat n_{1}} \qquad\label{bqb}
b_{+}b^{+}_{+} - \frac{1}{q}b^{+}_{+}b_{+} = q^{\hat n_{2}}
\end{equation}
The bozonization procedure of the above operators according to
Jannussis {\it et al.}\cite{JPB-78} yields the following expressions
for the $q$-deformed operators ($q$-bosons):
\begin{eqnarray}
                                                     \label{boson}
b_{-}=\sqrt{\frac{\bigl[\hat n_{1} +1\bigr]}{\hat n_{1} +1}}a_{-} \qquad
b^{+}_{-}=a^{+}_{-}\sqrt{\frac{\bigl[\hat n_{1} +1\bigr]}{\hat n_{1} +1}}\\
b_{+}=\sqrt{\frac{\bigl[\hat n_{2} +1\bigr]}{\hat n_{2} +1}}a_{+} \qquad
b^{+}_{+}=a^{+}_{+}\sqrt{\frac{\bigl[\hat n_{2} +1\bigr]}{\hat n_{2} +1}}
\end{eqnarray}
where
$\bigl[ x\bigr] = (q^{x}-q^{-x})/(q-q^{-1})$ and $a_{\pm}$ and
$a^{+}_{\pm}$ are given by (\ref{a-})-(\ref{a++}).
Due to these definitions, we can directly find that
\begin{eqnarray}
                                                     \label{bn}
b_{-}b^{+}_{-} = \bigl[ \hat n_{1} +1\bigr] \qquad
b^{+}_{-}b_{-} = \bigl[ \hat n_{1}   \bigr] \\
b_{+}b^{+}_{+} = \bigl[ \hat n_{2} +1\bigr] \qquad
b^{+}_{+}b_{+} = \bigl[ \hat n_{2}   \bigr]
\end{eqnarray}
Clearly, $b^{+}_{\pm}=(b_{\pm})^{\dagger}$ if $q \in R$ or $q \in S^{1}$.
The actions of the $q$-boson operators on the Fock space
${\cal F}_{1}\otimes{\cal F}_{2}$ with the basis
\begin{equation}
| n_{1}\, n_{2}\rangle =
\frac{(b^{+}_{-})^{n_{1}}(b^{+}_{+})^{n_{2}}}
{\sqrt{n_{1}!}\sqrt{n_{2}!}}| 0\, 0\rangle                    \label{basis}
\end{equation}
have the form
\begin{eqnarray}
                                                     \label{b-action}
b_{-}    | n_{1}\, n_{2}\rangle =
   \sqrt{\bigl[n_{1}\bigr]}  | n_{1}-1\, n_{2}\rangle \qquad
b^{+}_{-}| n_{1}\, n_{2}\rangle =
   \sqrt{\bigl[n_{1}+1\bigr]}| n_{1}+1\, n_{2}\rangle     \\
b_{+}    | n_{1}\, n_{2}\rangle =
   \sqrt{\bigl[n_{2}\bigr]}  | n_{1}\, n_{2}-1\rangle  \qquad
b^{+}_{+}| n_{1}\, n_{2}\rangle =
   \sqrt{\bigl[n_{1}+1\bigr]}| n_{1}\, n_{2}+1\rangle
\end{eqnarray}
In the following we consider the algebra implied by the generators
\begin{equation}
J_+ = b_{-}b^{+}_{+} \qquad
J_- = b_{+}b^{+}_{-}                     \label{J+-}
\end{equation}
It is a matter of straightforward calculations to find that
\begin{equation}
\bigl[J_{+},J_{-}\bigr] = \bigl[2J_{3}\bigr] \qquad
\bigl[J_{3},J_{\pm}\bigr] = \pm J_{\pm}      \label{J+J-}
\end{equation}
where
\begin{equation}
2J_{3} = \hat n_{1} - \hat n_{2}                     \label{J3}
\end{equation}
One can recognize that the above relations are standard quantum
algebra $su_{q}(2)$ commutation relations,
in the Kulish-Reshetikhin-Drinfeld-Jimbo
realization\cite{Biedenharn-89}-\cite{SK-92} according to which
$su_{q}(2)$ can be realized by two commuting sets of $q$-bosons
($q-$deformed version of the Jordan-Schwinger approach to angular
momentum). Hereafter, we write $su_{q}(2)$ to denote the
quantum algebra which is in fact $U_{q}(su(2))$.

 Comparing (\ref{J3}) with  (\ref{Wnn}) we see that the Wigner
operator for harmonic oscillator is just proportional to the
3-axis projection of the ($q$-deformed) spherical angular momentum
operator,
\begin{equation}
W_{-} = 2\hbar\omega J_{3}              \label{WJ}
\end{equation}

Indeed, in the $su(2)$ notations\cite{Macfarlane-89}
for basis vector $|n_{1}n_{2}\rangle$,
\begin{equation}
|j\, m\rangle = |n_{1}\, n_{2}\rangle \qquad         \label{jm-state}
j = \frac{1}{2}(n_{1}+n_{2}) \qquad
m = \frac{1}{2}(n_{1}-n_{2})
\end{equation}
the operators $J_{\pm}$ and $J_{3}$ act on
${\cal F}_{1}\otimes{\cal F}_{2}$ according to
\begin{eqnarray}
                                                     \label{J-action}
J_{-}|j\, m\rangle =
     \sqrt{\bigl[j+m\bigr]\bigl[j-m+1\bigr]}|j\, m-1\rangle \nonumber\\
J_{+}|j\, m\rangle =
     \sqrt{\bigl[j-m\bigr]\bigl[j+m-1\bigr]}|j\, m+1\rangle  \\
J_{3}|j\, m\rangle =
     m|j\, m\rangle                          \nonumber
\end{eqnarray}
 For a fixed value $2j \in Z$, vector $|j\ m\rangle$ span the irrep $(j)$
of the quantum algebra $su_{q}(2)$.
 We assume that $q$ is not root of unity.
 Acordingly, the charge operator $J = \frac{1}{2}(\hat{n_{1}}+\hat{n_{2}})$
commutes with $J_{\pm ,3}$, and $J|j\, m\rangle = j|j\, m\rangle$.

 The indication of the Wigner density operator $W_{+}$ may be seen from
the following.
 The basic fact\cite{SK-92}
is that the vector  $|n_{1}n_{2}\rangle \equiv |j\ m\rangle$
can be represented also as a basis vector $|k\ l\rangle$ for the irrep
beloning to the positive discrete series of
$su_{q}(1,1) \approx sp_{q}(2,R)$ with the hyperbolic angular momentum,
$k = \frac{1}{2}(n_{1}-n_{2}-1)= m-\frac{1}{2}$,
and 3-axis projection,
$l = \frac{1}{2}(n_{1}+n_{2}+1)= j+\frac{1}{2}$.
The generators of $su_{q}(1,1)$ are
\begin{equation}
K_{+} = b^{+}_{+}b^{+}_{-} \qquad
K_{-} = b_{+}b_{-}         \qquad
K_{3} = J + \frac{1}{2}
\end{equation}
Particularly, the 3-axis hyperbolic angular momentum operator $K_{3}$
acts due to
\begin{equation}
K_{3}|k\ l\rangle = l|k\ l\rangle
\end{equation}
Thus, the Wigner density operator
$W_+ = H(\Phi_{-}) + H(\Phi_{+}) = \hbar\omega (\hat n_{1}+\hat n_{2}+1)$
can be immediately identified with $K_{3}$,
\begin{equation}
W_{+} = 2\hbar\omega K_{3}              \label{WK}
\end{equation}

To summarize, we note that the harmonic ($q$-)oscillator in phase space
naturally arises to the Jordan-Schwinger approach
to ($q$-deformed) angular momentum, with the Wigner operator
$W_{-}$ ($W_{+}$) being
identified with the 3-axis spherical (hyperbolical) angular
momentum operator.

As a final remark, we notice that there are ways to give geometrical
interpretation of the quantum algebras and its representations.
Namely, one may follow the line of reasoning
by Fiore\cite{Fiore-93} and construct a realization of the
quantum algebra within $Diff(M_{q})$, where $M_{q}$ is
a $q$-deformed version of the ordinary manifold.
For example, in the context of the $q$-oscillator in phase space
it is highly interesting to find such a realization
for the algebra $su_{q}(1,1) \approx sp_{q}(2,R)$, which
is concerned the $q$-deformed phase space.

Also, there is a possibility\cite{Franco-93} to give
a geometric interpretation of the representations of $su_{q}(2)$
following the lines of the standard Borel-Weyl-Bott
theory\cite{Wallach-73,Fulton-91}.

\section{Conclusions}

 We studied the relation between the Bopp-Kubo formulation and
the Weyl-Wigner-Moyal calculus of quantum mechanics in phase space
which is found to arise from the fact that the Moyal product of
functions on phase space,
$f(\phi )*g(\phi )$,
can be rewritten equivalently as the product of functions
defined on the extended phase space,
$f(\Phi )g(\phi )$.

 From the non-commutative geometry point of view, the phase-space
formulation of quantum mechanics is an example of the theory
with non-commutative geometry. The non-commutative algebra ${\cal A}$
is the algebra of functions on phase space endowed with the
Moyal product. The right and left ${\cal A}$-modules are interchanged
by the modular conjugation ${\cal J}$ so that the space of sections
${\cal E} ={\cal A} \otimes {\cal A}/{\cal J}$, and there is
a kind of chiral symmetry due to the non-commutativity.

 Due to a similarity between the phase-space formulation of quantum
mechanics and Hamiltonian formulation of classical mechanics,
there is an attractive possibility to extend useful classical
notions and tools to quantum mechanics. An attempt is made
to formulate the quantum extension of the classical ergodicity
condition.

We studied one-dimensional harmonic ($q$-)oscillator in phase space.
The phase space has a $2D$-lattice structure, similar to the one
of the magnetic group periodicity for the $2D$-electron gas in
magnetic field.

 The Fock space for the oscillator is related to the double
Hilbert space of the Gelfand-Naimark-Segal construction in
accordance with the relation between the symbol calculus and
the GNS construction.

 For the $q$-oscillator, the Wigner operator $W_{-}$($W_{+}$)
is found to be proportional to the
3-axis spherical (hyperbolical) angular momentum operator of the
$q$-deformed algebra $su_{q}(2)$ ($su_{q}(1,1)\approx sp_{q}(2,R)$).

 The analysis of this paper is valid at fixed value of time. Its
reformulation in a form invariant under time evolution will be studied in a
future paper.

\end{document}